\documentclass[lettersize,journal]{IEEEtran}
\usepackage{amsmath,amsfonts}
\usepackage{algorithmic}
\usepackage{algorithm}
\usepackage{array}
\usepackage[caption=false]{subfig}
\usepackage{textcomp}
\usepackage{stfloats}
\usepackage{url}
\usepackage{verbatim}
\usepackage{graphicx}
\usepackage{cite}
\usepackage{color}
\usepackage{multirow}
\usepackage{multicol}
\usepackage{makecell}
\usepackage{hyperref}
\usepackage{booktabs}

\begin{document}

\title{Efficient Deployment of Deep MIMO Detection Using Learngene}

\author{Jinya~Zhang,~Jiajia~Guo,~Xiangyi~Li,~Chao-Kai~Wen, \IEEEmembership{\normalsize{Fellow,~IEEE}}, 

Xin Geng, \IEEEmembership{\normalsize{Senior Member,~IEEE}}, and Shi~Jin, \IEEEmembership{\normalsize{Fellow,~IEEE}}
        % <-this % stops a space
% <-this % stops a space
}

\maketitle
\begin{abstract} 
Deep learning (DL) has introduced a new paradigm in multiple-input multiple-output (MIMO) detection, balancing performance and complexity. However, the practical deployment of DL-based detectors is hindered by poor generalization, necessitating costly retraining for different devices and scenarios. To address this challenge, this paper presents a novel knowledge transfer technique, termed learngene, for the design of a DL-based MIMO detector and proposes an efficient deployment framework.
The proposed detector, SDNet, leverages zero-forcing detection outputs and least squares-estimated channel state information (CSI) as inputs. It is further optimized through a collective-individual paradigm to enhance knowledge transfer. In this paradigm, learngene, a reusable neural network (NN) segment, encapsulates detection meta-knowledge acquired from large-scale collective models trained by manufacturers. This segment can then be distributed to device-specific teams. By integrating learngene into different lightweight individual models, detection meta-knowledge is efficiently transferred across heterogeneous NNs, enabling adaptation to diverse devices and scenarios.
Simulation results demonstrate that the proposed scheme enhances performance, enables rapid adaptation, and ensures high scalability, with transferred parameters comprising only 10.8\% of the total model size.
\end{abstract}


\section{Introduction}
\IEEEPARstart{M}{ultiple}-input multiple-output (MIMO) technology has been integral to wireless standards from 4G to emerging 6G, significantly improving spectral efficiency and transmission performance \cite{ITU2022}. However, a persistent gap remains between practical MIMO performance and theoretical capacity limits \cite{yang2015fifty,albreem2019massive}. Traditional MIMO detection methods face notable challenges: while the maximum likelihood (ML) detector is theoretically optimal, its NP-hard complexity makes it impractical for large-scale arrays. Linear detectors, such as zero-forcing (ZF) and minimum mean squared error (MMSE), offer lower complexity but suffer from suboptimal performance and high computational overhead. Advanced techniques, including approximate message passing (AMP) and soft interference cancellation (SIC), rely on precise statistical knowledge of signal and interference characteristics, which is often difficult to obtain. These challenges underscore the need for innovative detection strategies.
 
The success of artificial intelligence (AI) in domains such as computer vision \cite{zhao2020exploring,tu2022maxim} and speech recognition \cite{radford2023robust} has catalyzed an AI-driven paradigm shift in wireless communications. As system complexity increases, AI-native communication systems have emerged \cite{o2017introduction}, and redesigning physical layer modules using deep learning (DL) has demonstrated significant efficacy \cite{guo2022overview,qin2019deep,yang2020deep,song2022benchmarking,ye2018power}. Notably, DL-based MIMO detectors have outperformed traditional approaches in various scenarios \cite{albreem2021deep,ye2018power,gao2018comnet,samuel2019learning,khani2020adaptive,goutay2020deep,ye2024purely}, leveraging neural networks (NNs) trained on extensive datasets to estimate transmitted signals from received signals.

Recent advancements have further improved DL-based MIMO detectors. For example, \cite{ye2018power} introduces a fully connected (FC) deep NN-based receiver that jointly optimizes channel estimation and signal detection, demonstrating DL’s potential for MIMO detection. In \cite{gao2018comnet}, a block-based architecture, ComNet, incorporates a channel estimation subnet and a signal detection subnet, both initialized with conventional solutions and refined through DL. Instead of stacking FC layers, \cite{samuel2019learning} proposes DetNet, which unfolds projected gradient descent iterations into a network using a linear transformation of the received signal vector. More recently, \cite{ye2024purely} presents ChannelNet, embedding the channel matrix within the NN as linear layers, enabling scalable detection for massive MIMO scenarios. 

Despite these advancements, key limitations persist. As highlighted in \cite{hu2023understanding}, DL-based detectors require extensive training datasets and carefully designed NN architectures. Furthermore, channel state information (CSI) is crucial for these detectors, with many approaches incorporating CSI as an explicit input. These factors introduce three major challenges. First, acquiring sufficient CSI for training is time-intensive and resource-demanding, often leading to data scarcity. Second, using CSI as an input increases NN complexity, hindering convergence in large-scale MIMO systems. Third, CSI statistics vary across environments (e.g., line-of-sight (LoS) vs. non-LoS (NLoS)  conditions, path loss, and delay), causing significant performance degradation when applying a trained NN to new scenarios, necessitating costly retraining and incurring high generalization errors.

Two potential solutions address these challenges. The first is knowledge transfer for DL-based detectors \cite{van2022transfer,huo2022intelligent,wang2023deep,shang2024design,uyoata2024transfer}, which leverages knowledge from prior domains or tasks \cite{pan2009survey} to improve learning efficiency. Knowledge transfer mitigates the need for extensive training data, accelerates model training, and reduces computational overhead, enhancing generalization and resource efficiency. However, existing knowledge transfer methods in MIMO detection often lack flexibility, struggle to accommodate devices with diverse computational constraints (e.g., smartphones, wearables), and require advanced AI expertise. The second approach involves training large-scale models on diverse datasets to improve generalization \cite{hoffmann2022training}. However, large models entail substantial training costs and prolonged inference times \cite{shen2023efficient}, making them impractical for resource-limited devices. Consequently, neither approach fully meets practical deployment requirements.

To address these limitations, learngene, a novel knowledge transfer framework inspired by biological inheritance, has been introduced \cite{wang2022learngene}. This approach enables efficient adaptation of small models by leveraging the generalization capability of large models. Specifically, \cite{wang2022learngene} proposes a collective-individual paradigm: a large-scale collective model is trained on sequential tasks, extracting meta-knowledge into lightweight NN segments---referred to as learngene---based on gradient analysis. Individual models inherit these learngene units and extend them flexibly, facilitating rapid convergence. Since learngene is computationally lightweight, it can be tailored to different devices. Recently, this method has been applied to CSI feedback, improving generalization while reducing training overhead for base station (BS) manufacturers \cite{li2024learngene}. This advancement underscores learngene’s potential in DL-based wireless communications.
 
To tackle the challenges of poor generalization, high training costs, and inflexibility in DL-based MIMO detection, we propose integrating learngene into DL-based detectors. The key contributions of this study are summarized as follows: 
\begin{itemize} 
\item \textbf{We analyze the generalization limitations of SDNet}, a DL-based MIMO detector, and investigate the correlation between symbol error rate (SER) degradation and CSI similarity. Through mathematical analysis, we establish that SDNet functions as a denoising NN, where noise introduced by ZF detector follows a CSI-dependent Gaussian distribution, limiting its generalization capability.

\item \textbf{We integrate learngene into SDNet to enhance generalization}. By training a large-scale collective model and extracting gradient-based learngene units, we enable lightweight individual model adaptation. Experiments demonstrate that learngene significantly improves detection accuracy and accelerates convergence, with the transferred parameters constituting only 10.8\% of the overall model. 

\item \textbf{We propose an efficient deployment framework} where detector manufacturers develop and distribute learngene units to device-specific teams. This approach facilitates efficient knowledge transfer and enables flexible customization, reducing deployment costs while maintaining high performance. 
\end{itemize}

The remainder of this paper is structured as follows. Section \ref{sec: System model and signal detection} introduces the system model and traditional MIMO detection algorithms. Section \ref{sec: Deep learning-based MIMO detection} details the DL-based MIMO detector design. Section \ref{sec: Meta-knowledge utilization} presents the integration of learngene into MIMO detection and its deployment strategy. Section \ref{sec: Simulation results} provides simulation results, and Section \ref{sec: Conclusion} concludes the paper. 
 
\textit{Notations:} Lowercase symbols denote scalars, bold lowercase symbols denote column vectors, and bold uppercase symbols denote matrices. The real and imaginary parts of complex matrices are denoted as $\Re(\cdot)$ and $\Im(\cdot)$, respectively. The symbols $(\cdot)^T$ and $(\cdot)^H$ represent the transpose and Hermitian of a matrix or vector, respectively. $(\cdot)^{-1}$ denotes the inverse of a matrix, $(\cdot)^{+}$ denotes the pseudo-inverse of a matrix, and $\text{Tr}(\cdot)$ denotes the trace operation. The notation $\|\cdot\|_2$ represents the 2-norm of a vector.

\section{System model and signal detection}
\label{sec: System model and signal detection}

\subsection{MIMO System}
We consider an uplink MIMO system where a BS with $N_\mathrm{r}$ antennas serves a single user equipment (UE) equipped with $N_\mathrm{t}$ antennas, where $N_\mathrm{r}>N_\mathrm{t}$. The channel matrix between the UE and BS is denoted as  
$\bar{\mathbf{H}} \in \mathbb{C}^{N_\mathrm{r} \times N_\mathrm{t}}$. Let $\bar{\mathbf{x}} \in \bar{\mathbb{S}}^{N_\mathrm{t}}$ be the transmitted symbol vector, where each element belongs to the constellation set $\bar{\mathbb{S}}$. We adopt quadrature phase shift keying (QPSK) as the modulation scheme. The received signal vector $\bar{\mathbf{y}} \in \mathbb{C}^{N_\mathrm{r}}$ is given by
\begin{equation}
    \bar{\mathbf{y}}=\bar{\mathbf{H}}\bar{\mathbf{x}}+\bar{\mathbf{n}},
    \label{eq: complex model}
\end{equation}
where $\bar{\mathbf{n}} \in \mathbb{C}^{N_\mathrm{r}}$ is the additive white Gaussian noise (AWGN) vector, whose elements follow independent, zero-mean complex normal distributions with variance $\mathbf{\sigma}_\mathrm{n}^2$.

To facilitate NN processing, we convert the complex model into an equivalent real-valued representation. Let $\mathbf{x} = [\Re(\bar{\mathbf{x}})^T,\Im(\bar{\mathbf{x}})^T ]^T \in \mathbb{S}^{2N_\mathrm{t}}$, 
$\mathbf{y} = [\Re(\bar{\mathbf{y}})^T, \Im(\bar{\mathbf{y}})^T ]^T \in \mathbb{R}^{2N_\mathrm{r}}$, and 
$\mathbf{n} = [\Re(\bar{\mathbf{n}})^T,\Im(\bar{\mathbf{n}})^T ]^T \in \mathbb{R}^{2N_\mathrm{r}}$, where $\mathbb{S}=\Re(\bar{\mathbb{S}})=\Im(\bar{\mathbb{S}})$. The real-valued equivalent of the channel matrix is defined as 
\begin{equation}
    \mathbf{H} = \begin{bmatrix} 
        \Re(\bar{\mathbf{H}}) & -\Im(\bar{\mathbf{H}}) \\ 
        \Im(\bar{\mathbf{H}}) & \Re(\bar{\mathbf{H}}) 
    \end{bmatrix} 
    \in \mathbb{R}^{2N_\mathrm{r} \times 2N_\mathrm{t}}.
    \label{eq: real H}
\end{equation}
Thus, the real-valued representation of (\ref{eq: complex model}) becomes
\begin{equation} \label{eq:y=Hx+n}
    \mathbf{y}=\mathbf{H}\mathbf{x}+\mathbf{n}.
\end{equation}

MIMO detection aims to recover $\mathbf{x}$ from the received signal $\mathbf{y}$ and the estimated channel matrix $\mathbf{H}$ with minimal error. In this work, we assume the channel matrix is estimated using pilot-based channel estimation. Let $\bar{\mathbf{X}}_\mathrm{p} \in \bar{\mathbb{S}}^{N_\mathrm{t} \times N_\mathrm{p}}$ denote the pilot symbol matrix, and $\bar{\mathbf{N}}_\mathrm{p} \in \mathbb{C}^{N_\mathrm{r} \times N_\mathrm{p}}$ represent the AWGN matrix, where $N_\mathrm{p}$ is the number of pilot symbols. The received pilot signal matrix $\bar{\mathbf{Y}}_\mathrm{p} \in \mathbb{C}^{N_\mathrm{r} \times N_\mathrm{p}}$ is 
\begin{equation}
    \bar{\mathbf{Y}}_\mathrm{p}=\bar{\mathbf{H}}\bar{\mathbf{X}}_\mathrm{p}+\bar{\mathbf{N}}_\mathrm{p}.
    \label{eq: pilot complex model}
\end{equation}
The least squares (LS) channel estimate $\hat{\bar{\mathbf{H}}}_\mathrm{LS} \in \mathbb{C}^{N_\mathrm{r} \times N_\mathrm{t}}$  is given by
\begin{equation}
    \hat{\bar{\mathbf{H}}}_\mathrm{LS} = \bar{\mathbf{Y}}_\mathrm{p}({\bar{\mathbf{X}}_\mathrm{p}}^H\bar{\mathbf{X}}_\mathrm{p})^{-1}{\bar{\mathbf{X}}_\mathrm{p}}^H.
    \label{eq: LS estimation}
\end{equation}
Using (\ref{eq: real H}), the real-valued LS channel estimate $\hat{\mathbf{H}}_\mathrm{LS} \in \mathbb{R}^{2N_\mathrm{r} \times 2N_\mathrm{t}}$ is obtained accordingly. 
The ML estimate of the transmitted symbol vector is formulated as
\begin{equation}
    \hat{\mathbf{x}}_\mathrm{ML} = \arg\min_{\mathbf{x} \in \mathbb{S}^{2N_t}} \| \mathbf{y} - \hat{\mathbf{H}}_\mathrm{LS} \mathbf{x} \|_2^2.
    \label{eq: MLdetect}
\end{equation}
Although ML detection is optimal, solving (\ref{eq: MLdetect}) is NP-hard due to the discrete nature of $\mathbb{S}$, making it computationally prohibitive for large-scale MIMO systems.

\subsection{Traditional Linear MIMO Detectors}
Linear MIMO detectors are widely used due to their low computational complexity. These methods mitigate interference by applying a linear transformation to the received signal, aiming to reverse the channel's effects and isolate the transmitted signals. The general detection strategy of a linear detector is given by
\begin{equation}
    \hat{\mathbf{x}} = \mathbf{A}\mathbf{y},
\end{equation}
where $\mathbf{A}$  is the equalization matrix that counteracts the influence of the channel. Two commonly used linear detectors are ZF   and MMSE  detectors.
 
\textbf{ZF detector:} The ZF detector applies the Moore-Penrose pseudo-inverse of the estimated channel matrix $\hat{\mathbf{H}}_\mathrm{LS}$ as the equalization matrix
\begin{equation}
    \mathbf{A}_\mathrm{ZF}=\mathbf{\hat{H}}_\mathrm{LS}^+=({\hat{\mathbf{H}}_\mathrm{LS}}^T\hat{\mathbf{H}}_\mathrm{LS})^{-1}{\hat{\mathbf{H}}_\mathrm{LS}}^T.
    \label{eq: wzf}
\end{equation}
The estimated transmitted symbol vector is then obtained as
\begin{equation}
    \hat{\mathbf{x}}_\mathrm{ZF}=\mathbf{A}_\mathrm{ZF}\mathbf{y}=({\hat{\mathbf{H}}_\mathrm{LS}}^T\hat{\mathbf{H}}_\mathrm{LS})^{-1}{\hat{\mathbf{H}}_\mathrm{LS}}^T\mathbf{y}.
    \label{eq: xzf}
\end{equation}
However, the ZF detector does not account for the impact of AWGN, leading to high sensitivity to noise, which can significantly degrade performance, particularly in low signal-to-noise ratio (SNR) conditions.

\textbf{MMSE detector:} The MMSE detector improves upon ZF by incorporating noise variance into the equalization process. The MMSE equalization matrix is given by
\begin{equation}
    \mathbf{A}_\mathrm{MMSE}=({\hat{\mathbf{H}}_\mathrm{LS}}^T\hat{\mathbf{H}}_\mathrm{LS}+\sigma^2_\mathrm{n}\mathbf{I}_{2N_\mathrm{t}})^{-1}{\hat{\mathbf{H}}_\mathrm{LS}}^T,
\end{equation}
where $\mathbf{I}_{2N_\mathrm{t}}$ is the identity matrix. The estimated transmitted symbol vector is then computed as
\begin{equation}
    \hat{\mathbf{x}}_\mathrm{MMSE}=\mathbf{A}_\mathrm{MMSE}\mathbf{y}=({\hat{\mathbf{H}}_\mathrm{LS}}^T\hat{\mathbf{H}}_\mathrm{LS}+\sigma^2_\mathrm{n}\mathbf{I}_{2N_\mathrm{t}})^{-1}{\hat{\mathbf{H}}_\mathrm{LS}}^T\mathbf{y}.
\end{equation}
Compared to the ZF detector, the MMSE detector provides improved performance by considering the effect of noise, leading to enhanced robustness, especially in low SNR regimes.

\section{DL-based MIMO Detection}
\label{sec: Deep learning-based MIMO detection}

In this section, we introduce SDNet, a DL-based MIMO detection framework that leverages ZF detection results and estimated CSI as inputs. First, we describe the overall framework, followed by a detailed explanation of SDNet’s architecture. Lastly, we provide a mathematical analysis of its learning process.

\subsection{Framework}
\begin{figure*}[!t]
    \centering
    \includegraphics[width=0.85\textwidth]{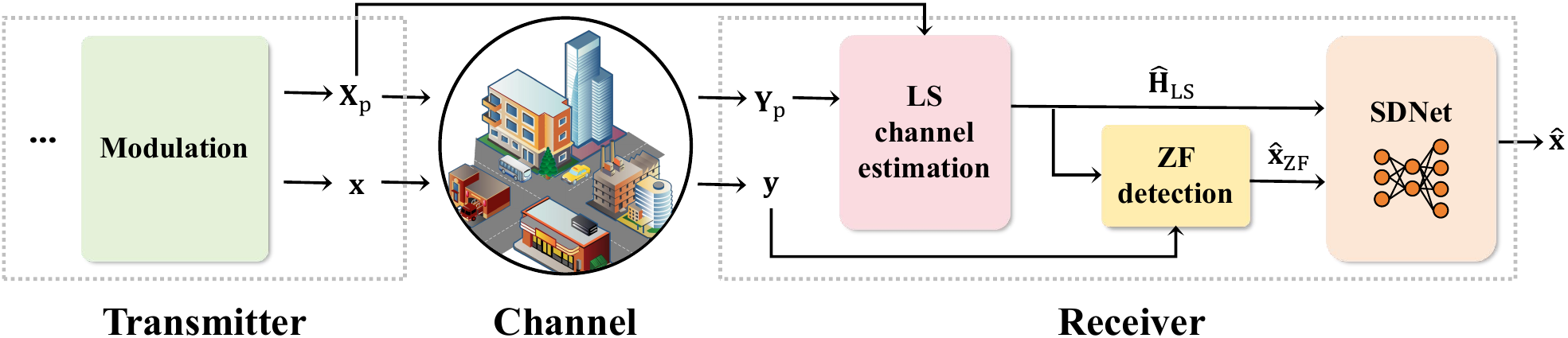}
    \caption{Framework of the DL-based MIMO detection system.} 
    \label{fig: detectionNN}
    \vspace{0.5cm}
\end{figure*}

The proposed framework, illustrated in Fig. \ref{fig: detectionNN}, is inspired by \cite{gao2018comnet}. It operates in a MIMO system where both the pilot symbol matrix $\mathbf{X}_\mathrm{p}$ and the data symbol vector $\mathbf{x}$ are transmitted. Consequently, the receiver obtains the pilot signal matrix $\mathbf{Y}_\mathrm{p}$ and the received data vector $\mathbf{y}$,  which are jointly processed to reconstruct the original data. Our framework integrates a traditional channel estimation module with a DL-based detection module, augmented by the ZF algorithm.  

\paragraph{Channel Estimation Module}
In each coherence time, the LS channel estimation algorithm computes the estimated CSI $\hat{\mathbf{H}}_\mathrm{LS}$ using pilot transmissions, as described in (\ref{eq: real H}), (\ref{eq: pilot complex model}), and (\ref{eq: LS estimation}). This estimated CSI is utilized in the subsequent signal detection process.

\paragraph{Signal Detection Module}
The signal detection process comprises two stages:
\begin{itemize}
    \item \textbf{Coarse detection:} A low-complexity traditional detection step generates an initial estimate of the transmitted signal.
    \item \textbf{DL-based refinement:} An NN, referred to as SDNet, further refines the coarse detection result to improve accuracy.
\end{itemize}

First, the ZF algorithm produces a coarse estimate $\hat{\mathbf{x}}_\mathrm{ZF}$, computed as per (\ref{eq: wzf}) and (\ref{eq: xzf}). Then, both the ZF detection result $\hat{\mathbf{x}}_\mathrm{ZF}$ and the estimated CSI  $\hat{\mathbf{H}}_\mathrm{LS}$ are fed into SDNet to generate a more precise symbol vector $\hat{\mathbf{x}}$
\begin{equation}
    \hat{\mathbf{x}}=f(\{\hat{\mathbf{x}}_\mathrm{ZF},\hat{\mathbf{H}}_\mathrm{LS}\};\mathbf{\Theta}),
\end{equation}
where $f(\cdot)$ represents the detection function, and $\mathbf{\Theta}$ denotes the NN parameters.

As highlighted in \cite{gao2018comnet}, integrating conventional ZF detection into a DL-based framework significantly improves performance. The ZF algorithm provides an initial estimate, allowing the NN to learn residual corrections and refine $\hat{\mathbf{x}}_\mathrm{ZF}$ during training. Although $\hat{\mathbf{x}}_\mathrm{ZF}$ implicitly contains channel knowledge, incorporating $\hat{\mathbf{H}}_\mathrm{LS}$ as an explicit input enables SDNet to learn channel features more effectively. The learngene extraction and expansion process will be built upon SDNet, as detailed later in Section \ref{sec: Meta-knowledge utilization}.

\begin{figure*}[!t]
    \centering
    \includegraphics[width=0.9\textwidth]{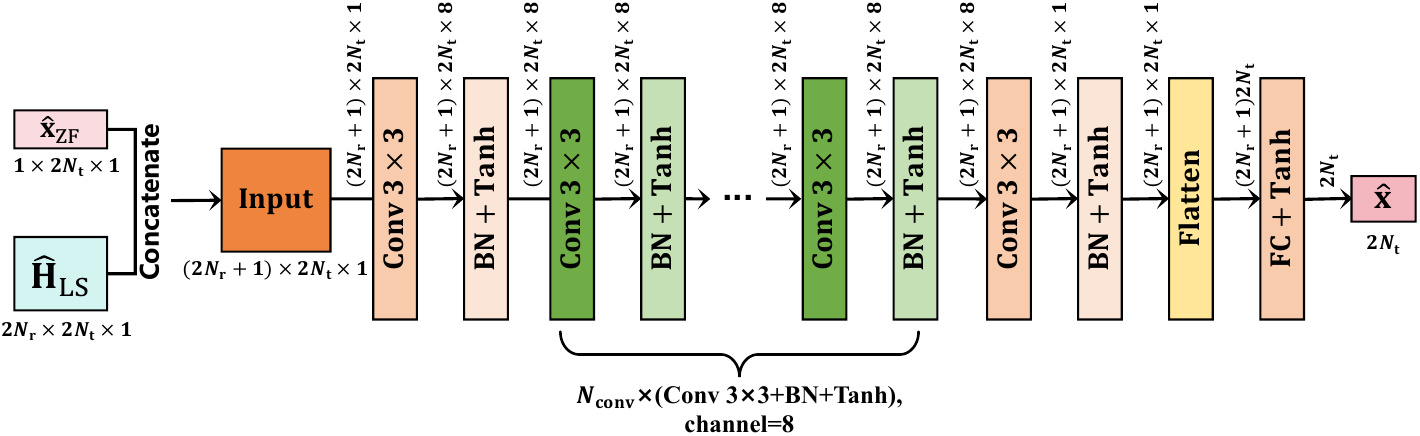}
    \caption{NN architecture of SDNet.}
    \label{fig: NNarchitecture}
\end{figure*}

\subsection{NN Architecture}
The overall architecture of SDNet is illustrated in Fig. \ref{fig: NNarchitecture}. The NN takes two inputs: the ZF-detected symbol vector $\hat{\mathbf{x}}_\mathrm{ZF}$ and the estimated CSI $\hat{\mathbf{H}}_\mathrm{LS}$. These inputs are first reshaped and concatenated into a single input matrix of dimension $(2N_\mathrm{r}+1) \times 2N_\mathrm{t} \times 1$, which is then processed by SDNet. 

SDNet consists of $N_\mathrm{conv}+2$ convolutional (Conv) layers, batch normalization (BN) layers, and an FC layer. The depth parameter $N_\mathrm{conv}$ determines the network complexity. The first $N_\mathrm{conv}+1$ Conv layers use a $3 \times 3$  kernel with 8 channels, while the final Conv layer uses a $3 \times 3$ kernel with 1 channel. Each Conv layer is followed by a BN layer and a Tanh activation function. After the ($N_\mathrm{conv}+2$)-th Tanh activation, a flattening layer reshapes the tensor into size $(2N_\mathrm{r}+1) 2N_\mathrm{t}$. Finally, an FC layer with a Tanh activation function produces the refined symbol vector $\hat{\mathbf{x}}$.

Training SDNet involves optimizing a loss function using a large-scale dataset containing training samples $\{ \hat{\mathbf{x}}_\mathrm{ZF} (n),  \hat{\mathbf{H}}_\mathrm{LS} (n) \}$, where  $(n)$ indexes the samples. The objective is to minimize the discrepancy between the predicted output $\hat{\mathbf{x}}(n)$ and the ground truth transmitted symbol vector $\mathbf{x}(n)$, enabling the model to learn an effective mapping from inputs to outputs. SDNet employs the mean squared error (MSE) loss function, formulated as 
\begin{equation}
    \min_{\mathbf{\Theta}} 
    \frac{1}{N}\sum^{N}_{n=1}\|\hat{\mathbf{x}} (n)-\mathbf{x}(n)\|_2^2,
\end{equation}
where $N$ is the total number of training samples.

\subsection{Analysis of SDNet}
\label{subsec: Analysis of SDNet}
The primary function of SDNet can be interpreted as denoising the coarse ZF detection output. To gain deeper insights into the network's learning process, we analyze the mathematical relationship between the input $\hat{\mathbf{x}}_\mathrm{ZF}$ and output $\hat{\mathbf{x}}$ through mathematical analysis. The subsequent derivation is conducted in the complex domain. 

From (\ref{eq:y=Hx+n}) and (\ref{eq: xzf}), the ZF detection output can be expressed as
\begin{equation} 
    \hat{\bar{\mathbf{x}}}_\mathrm{ZF} 
    =(\hat{\bar{\mathbf{H}}}_\mathrm{LS}^H\hat{\bar{\mathbf{H}}}_\mathrm{LS})^{-1}\hat{\bar{\mathbf{H}}}_\mathrm{LS}^H\bar{\mathbf{H}}\bar{\mathbf{x}}+(\hat{\bar{\mathbf{H}}}_\mathrm{LS}^H\hat{\bar{\mathbf{H}}}_\mathrm{LS})^{-1}\hat{\bar{\mathbf{H}}}_\mathrm{LS}^H\bar{\mathbf{n}}.
    \label{eq: xzf1}
\end{equation}
Assuming perfect CSI, i.e.,  $\hat{\bar{\mathbf{H}}}_\mathrm{LS}=\bar{\mathbf{H}}$, (\ref{eq: xzf1}) simplifies to 
\begin{align}
    \hat{\bar{\mathbf{x}}}_\mathrm{ZF}
    &\approx \bar{\mathbf{x}}+(\bar{\mathbf{H}}^H\bar{\mathbf{H}})^{-1}\bar{\mathbf{H}}^H\bar{\mathbf{n}} \notag \\
    &=\bar{\mathbf{x}}+\bar{\mathbf{n}}_\mathrm{ZF}.
    \label{eq: xzf2}
\end{align}
Here, $\bar{\mathbf{n}}_\mathrm{ZF}$ represents an error component introduced by ZF detection, which follows a Gaussian distribution. The key goal of SDNet is to suppress $\bar{\mathbf{n}}_\mathrm{ZF}$, thereby refining the detection accuracy. To analyze the statistical properties of the error component, the covariance matrix of $\bar{\mathbf{n}}_\mathrm{ZF}$ is given by
\begin{equation}
    \mathbb{E}\{\bar{\mathbf{n}}_\mathrm{ZF}\bar{\mathbf{n}}_\mathrm{ZF}^H\}=\sigma_\mathrm{n}^2(\bar{\mathbf{H}}^H\bar{\mathbf{H}})^{-1}.
    \label{eq: covariance}
\end{equation}
In practical MIMO systems, due to the inherent correlation in the channel matrix $\bar{\mathbf{H}}$, its columns are typically non-orthogonal. As a result, $(\bar{\mathbf{H}}^H\bar{\mathbf{H}})^{-1}$ is typically non-diagonal. Consequently, $\bar{\mathbf{n}}_\mathrm{ZF}$ exhibit colored noise characteristics. Applying singular value decomposition (SVD) to the channel matrix $\bar{\mathbf{H}}$, expressed as $\bar{\mathbf{H}} = \mathbf{U}\mathbf{\Sigma}\mathbf{V}^H$, the aggregate variance of $\bar{\mathbf{n}}_\mathrm{ZF}$ can be characterized as
\begin{equation}
    \mathbb{E}\{\|\bar{\mathbf{n}}_\mathrm{ZF}\|_2^2\} =\sigma_\mathrm{n}^2\text{Tr}(\mathbf{\Sigma}^{-2}).
    \label{eq: variance}
\end{equation}

Equations (\ref{eq: covariance}) and (\ref{eq: variance}) explicitly illustrates the dependency of $\bar{\mathbf{n}}_\mathrm{ZF}$ on $\bar{\mathbf{H}}^H\bar{\mathbf{H}}$ and its eigenvalues. Since SDNet is designed to suppress $\bar{\mathbf{n}}_\mathrm{ZF}$, its performance is inherently tied to the channel conditions.  
A well-trained SDNet can effectively suppress $\bar{\mathbf{n}}_\mathrm{ZF}$ and improve detection accuracy. However, variations in the scatterer distribution across different wireless transmission environments alter the statistical properties of the channel matrix, leading to potential generalization issues.
Specifically, since SDNet is trained on a particular dataset, its performance may degrade in new environments with different channel statistics. This highlights the necessity for adaptive techniques such as meta-learning or knowledge transfer, discussed in the next.

\section{Meta-Knowledge Utilization with Learngene}
\label{sec: Meta-knowledge utilization}

To mitigate the high training costs associated with developing models with strong generalization capabilities, this study integrates learngene \cite{wang2022learngene} into SDNet. 

\subsection{Motivation}
\begin{table*}[!t]
\centering
\setlength{\tabcolsep}{8pt}
\renewcommand{\arraystretch}{1.7}
\caption{The SER performance of the SDNet when training and testing datasets match/mismatch.}
\label{tab: generalization}
\begin{tabular}{|cc|ccccc|}
\hline
\multicolumn{1}{|c|}{\multirow{2}{*}{\textbf{Training datasets}}} & \multirow{2}{*}{$N_\mathrm{conv}$} &  \multicolumn{5}{c|}{\textbf{Testing datasets}}  \\ 
\cline{3-7}
\multicolumn{1}{|c|}{}  &  & \multicolumn{1}{c|}{Task B} & \multicolumn{1}{c|}{Task C}  & \multicolumn{1}{c|}{Task D}  & \multicolumn{1}{c|}{Task E}  & Task F \\
\hline
\multicolumn{1}{|c|}{\makecell{The same as testing task}} &  \multirow{2}{*}{12}  & \multicolumn{1}{c|}{2.01e-02} & \multicolumn{1}{c|}{1.01e-04} &\multicolumn{1}{c|}{1.33e-02}  & \multicolumn{1}{c|}{2.30e-02}  & 5.87e-04 \\
\cline{1-1}
\cline{3-7}
\multicolumn{1}{|c|}{Task A} &   & \multicolumn{1}{c|}{3.00e-02} & \multicolumn{1}{c|}{1.14e-03} & \multicolumn{1}{c|}{1.81e-02}  & \multicolumn{1}{c|}{2.99e-02}  & 6.58e-04 \\
\hline
\multicolumn{2}{|c|}{Traditional ZF} & \multicolumn{1}{c|}{1.26e-01} & \multicolumn{1}{c|}{9.64e-02} & \multicolumn{1}{c|}{1.21e-01}  & \multicolumn{1}{c|}{1.25e-01}  & 6.48e-02   \\
\cline{1-7}
\multicolumn{2}{|c|}{Traditional MMSE} & \multicolumn{1}{c|}{9.82e-02}  & \multicolumn{1}{c|}{7.82e-02}  & \multicolumn{1}{c|}{1.17e-01}  & \multicolumn{1}{c|}{9.92e-02}  & 5.02e-02  \\
\hline
\end{tabular}
\end{table*}

\begin{figure}[!t]
    \centering
    \subfloat[Average Euclidean distances among datasets A to F.]{\includegraphics[width=0.4\textwidth]{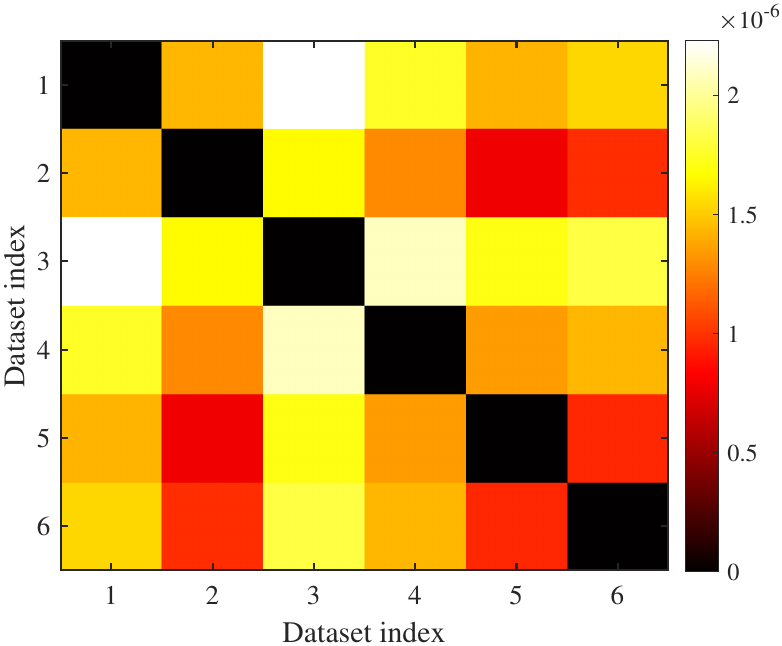}
    \label{fig: hotmap}}
    \hfil
    \subfloat[PCC between CSI dataset similarity and SER generalization error across the six datasets (A–F).]{\includegraphics[width=0.4\textwidth]{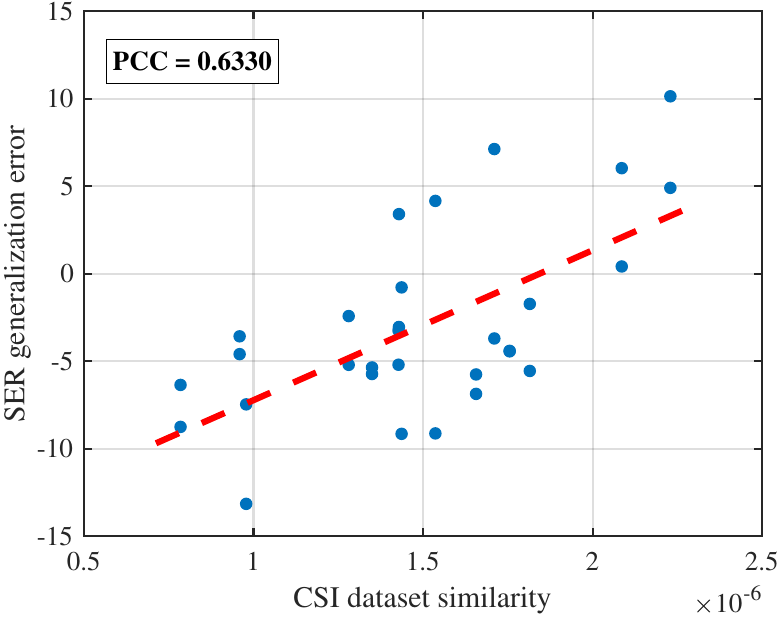}
    \label{fig: PCC}}
    \caption{Correlation analysis between SER degradation and CSI similarity.}
\end{figure}

While DL-based detectors achieve superior performance, they often suffer from limited generalization in dynamic wireless environments. To evaluate the generalization capability of SDNet, its performance is assessed across multiple tasks using models trained on different datasets. The simulation settings are detailed in Section \ref{sec: Simulation results}.
Table \ref{tab: generalization} presents the SER of SDNet under matched and mismatched training and testing datasets. As observed, mismatched datasets lead to higher SER, indicating reduced generalization. To quantify this effect, the Pearson correlation coefficient (PCC) is employed to analyze the correlation between CSI dataset similarity and SER generalization error \cite{morais2024dataset}. Dataset similarity is measured using average Euclidean distances among datasets A to F, as shown in Fig. \ref{fig: hotmap}. The SER generalization error is defined as
\begin{equation}
    \mathrm{SER}_{mn}=10 \log_{10}  \left| \frac{\mathrm{SER}_{m,n}-\mathrm{SER}_n}{\mathrm{SER}_n}\right| ,
\end{equation}
where $\mathrm{SER}_n$ represents the SER when training and testing occur on the same task $n$, while $\mathrm{SER}_{m,n}$ denotes the SER when training is performed on task $m$ and testing on task $n$. 
Simulation results indicate that the PCC between CSI dataset similarity and SER generalization error is 0.6330, suggesting a strong linear correlation, as depicted in Fig. \ref{fig: PCC}. These findings highlight the impact of dataset discrepancies on SDNet performance, revealing varying degrees of SER degradation when deployed in unseen environments. The need for retraining in new conditions imposes significant costs, underscoring the challenge of enhancing SDNet's generalization capability.

\subsection{Knowledge Transfer}
Knowledge transfer \cite{pan2009survey} enables efficient training of DL-based detectors by reducing computational overhead while maintaining robustness. Various studies have explored advanced knowledge transfer techniques in this domain.

\subsubsection{Transfer Learning}
Transfer learning, particularly the pre-training and fine-tuning strategy, involves training a model on a large-scale dataset and subsequently fine-tuning it for specific target tasks. This approach facilitates knowledge transfer from the source domain to the target domain, reducing training costs and improving generalization \cite{pan2009survey}.  For instance, \cite{van2022transfer} proposes a transfer learning framework for channel estimation and signal detection across different channel configurations, demonstrating improved training efficiency and detection accuracy. However, pre-trained models are inherently optimized for the source task, and low similarity between source and target tasks can lead to negative transfer. Additionally, the fine-tuned model’s scale is constrained by the pre-trained model, limiting customization across devices with varying computational capacities. In contrast, learngene extracts common knowledge from sequential learning, enabling flexible model design through a lightweight learngene unit.

\subsubsection{Meta Learning}
Meta learning, or learning-to-learn, distills knowledge from multiple related tasks to enhance the adaptability of NNs to new target tasks \cite{hospedales2021meta}. It is commonly categorized into metric-based, model-based, and optimization-based approaches, all designed to improve learning efficiency. In \cite{huo2022intelligent}, meta learning is applied to adaptive $K$-best-based MIMO detection, where the value of $K$ is determined by a fitting function whose coefficients are learned through meta learning. This enables the model to adaptively adjust $K$ while reducing complexity. However, meta learning typically requires batched task processing rather than sequential learning, and its optimization process demands substantial AI expertise. In contrast, learngene accumulates experience through sequential task training and employs parameter transfer to facilitate meta-knowledge inheritance, leveraging its structure as an NN segment.
 
\subsubsection{Multi-Task Learning}
Multi-task learning (MTL) simultaneously optimizes multiple related tasks to extract common knowledge, enhancing learning efficiency for a predefined set of known tasks \cite{zhang2021multi}. MTL is generally classified into: 
\begin{itemize} 
\item Hard parameter sharing, where lower-layer parameters are shared while task-specific parameters exist in upper layers. 
\item Soft parameter sharing, where task-specific models maintain separate parameters but are regularized to encourage similarity. 
\end{itemize}
MTL offers advantages such as reduced model complexity, lower storage requirements, and improved generalization. In \cite{sagduyu2023multi}, an MTL framework is introduced for semantic communication systems, where different decoders are trained for distinct tasks. However, MTL is inherently limited to a fixed set of known tasks and lacks adaptability to new, unseen tasks. In contrast, learngene distills meta-knowledge from sequential training, enabling efficient adaptation to new tasks and significantly enhancing generalization. 

In practical signal detection, while the aforementioned techniques reduce training costs and improve generalization from different perspectives, they remain constrained by limited adaptability to heterogeneous hardware, reliance on extensive AI expertise, or an inability to continuously accumulate knowledge from sequential tasks for application to unseen scenarios. These limitations pose significant challenges for deploying DL-based detectors across diverse communication environments. Therefore, an alternative knowledge transfer strategy is required to enable efficient accumulation and transfer of essential detection knowledge while allowing flexible adaptation. A promising solution is leveraging the superior reasoning and generalization capabilities of large-scale models to guide the learning process of lightweight models, i.e., applying learngene to DL-based detectors. 

\subsection{The Collective-Individual Paradigm}
\begin{figure*}[!t]
    \centering
    \includegraphics[width=1\textwidth]{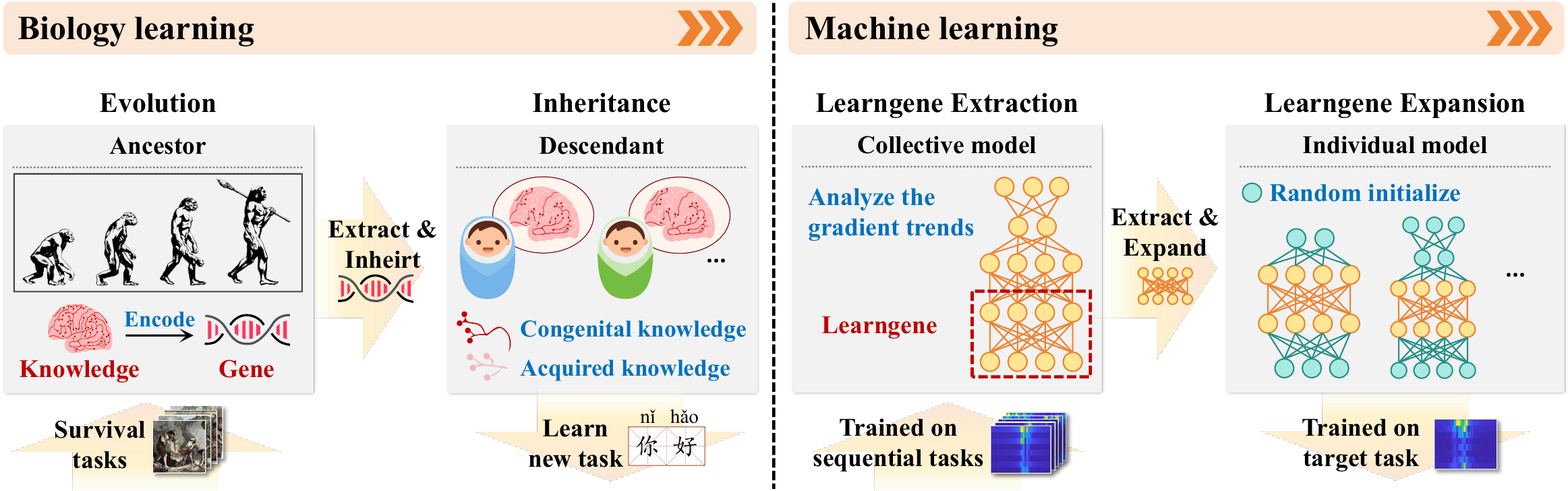}
    \caption{Illustration of knowledge transfer in individual models (right) inspired by genetic inheritance in biological learning (left).}
    \label{fig: biology and machine}
\end{figure*}

To enhance knowledge transfer and address the aforementioned challenges, we first examine how both biological and artificial systems acquire knowledge. While machines and animals learn through experience, their mechanisms differ fundamentally. In machines, knowledge is encoded within NNs, which rely on large-scale training datasets and carefully designed architectures. These networks are typically initialized randomly before training, meaning they begin without prior knowledge.
In contrast, as shown in the left half of Fig. \ref{fig: biology and machine}, biological systems retain and transmit knowledge through genetic inheritance. This enables offspring to inherit accumulated learning from their ancestors, facilitating innate behaviors and accelerating skill acquisition through environmental interactions. Unlike NNs, biological brains are not randomly initialized; instead, they leverage evolutionary learning, enhancing adaptability to new tasks. This evolutionary advantage offers valuable insights for improving machine learning processes.

To leverage these insights, Wang et al. \cite{wang2022learngene} propose a collective-individual paradigm comprising two key components: collective models and individual models. In this framework, the collective model continuously learns from sequential tasks, accumulating meta-knowledge analogous to evolutionary adaptation in biological systems. This knowledge is encapsulated in a module termed learngene, which is then used to initialize lightweight individual models. By inheriting meta-knowledge, individual models can rapidly adapt to new tasks, significantly reducing computational overhead. Notably, this process ensures data privacy, as there is no direct data exchange between collective and individual models.

In the context of signal detection, variations in deployment scenarios can degrade the performance of SDNet. Training high-performance detectors across all scenarios is computationally expensive, requiring substantial resources, extended training durations, and large-scale datasets. To mitigate these challenges, the collective-individual framework applies learngene to initialize SDNet for a given target task, optimizing the training process and significantly reducing computational costs. The detailed implementation of learngene and model training strategies is discussed in Sections \ref{subsec: Learngene} and \ref{subsec: Training strategies}.

\subsection{Learngene Extraction and Expansion}
\label{subsec: Learngene}

Learngene serves as the carrier of meta-knowledge, facilitating knowledge transfer across heterogeneous NNs. It consists of well-trained NN layers and is integrated into the collective-individual framework through two key phases: \textbf{learngene extraction} and \textbf{learngene expansion}, as illustrated in the right half of Fig. \ref{fig: biology and machine}.

During the learngene extraction phase, the collective model undergoes continuous training to accumulate knowledge, which can be categorized into task-specific knowledge and detection meta-knowledge. Since only meta-knowledge is essential for learning new tasks, the extraction process involves analyzing gradient trends across each layer of the collective model to gain insights into the training dynamics. If a layer’s parameters change significantly over sequential tasks, it suggests that its gradient remains large and has not yet converged. This indicates that the layer is highly sensitive to scenario variations, making it more likely to store task-specific knowledge. Conversely, if the parameters of a layer gradually stabilize and its gradient approaches zero from an initially high value, the layer is less sensitive to scenario changes and is more likely to contain meta-knowledge. Layers with this characteristic are selected as learngene, ensuring the extracted knowledge is transferable across different tasks.

\begin{figure}[!t]
    \centering
    \includegraphics[width=0.48\textwidth]{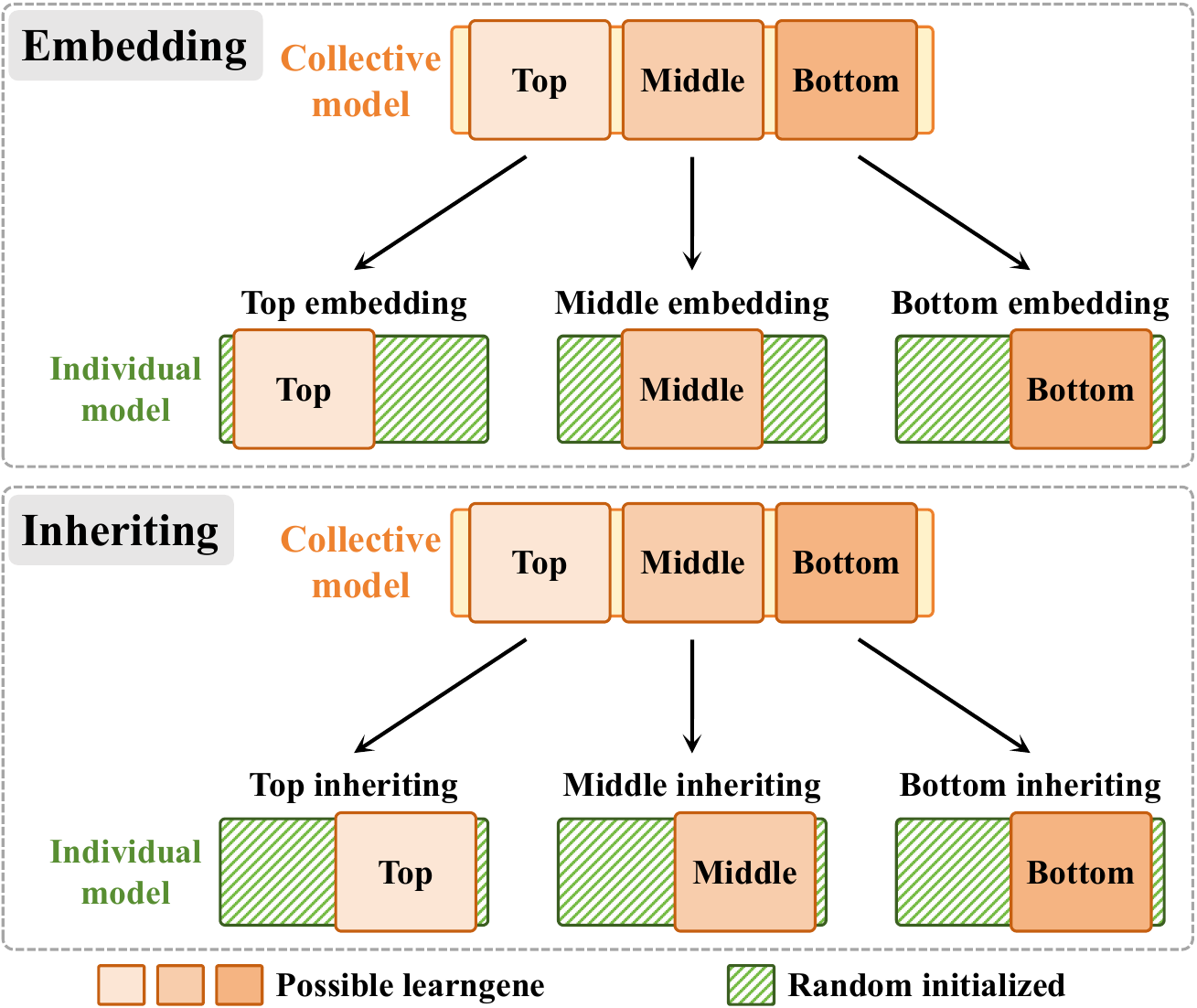}
    \caption{Expansion strategies for individual models.}
    \label{fig: expansion manners}
\end{figure}

Following extraction, the learngene expansion phase enhances individual models by integrating the extracted learngene. In this process, learngene initializes specific layers of the individual model, while the remaining layers are randomly initialized. The expansion follows two primary strategies: embedding, where learngene is directly incorporated into the individual model, and inheriting, where the knowledge from the collective model is transferred to the bottom of the individual model. Depending on the positioning of learngene within the collective model, these strategies can be further categorized into six subtypes: top, middle, and bottom embedding/inheriting, as shown in Fig. \ref{fig: expansion manners}. The positioning of learngene within the model significantly impacts knowledge transfer efficiency, as different layers contribute differently to model adaptation.
The effectiveness of knowledge transfer relies on both the correct extraction of learngene and the selection of an appropriate expansion strategy. By leveraging learngene, individual models inherit meta-knowledge accumulated by the collective model, allowing them to quickly adapt to target tasks while reducing computational overhead.

\subsection{Training Strategies}
\label{subsec: Training strategies}
As discussed in Section \ref{subsec: Learngene}, the initial phase of training involves extracting learngene through sequential training. To evaluate the importance of each layer, gradient significance is introduced as a metric to analyze gradient variations across sequential tasks \cite{wang2022learngene}. Given a set of sequential tasks denoted as $\mathcal{D} = \{ \mathcal{D}_1,\mathcal{D}_2,\dots,\mathcal{D}_K\}$, the collective model's gradient magnitude for each layer across different tasks is formulated as
\begin{equation}  
    g_{i,k}^l = \left|  \nabla_{i} \mathcal{L}(\mathbf{\Theta}^l_\mathrm{col}; \mathcal{D}_k) \right|,
\end{equation}
where $\mathcal{L}$ is the MSE loss function, and $\mathbf{\Theta}_\mathrm{col}$ represents the collective model parameters. Here, $k \in \{1,\dots,K\}$ is the task index in sequential training, $l \in \{1,\dots, L\}$  is the layer index, and $i \in \{1,\dots,\left| \mathbf{\Theta}_\mathrm{col}^l \right| \}$ denotes the parameter index within each layer. 

During sequential training, if the gradient of a layer fluctuates significantly before gradually stabilizing to zero, i.e., $g_{i,k}^l$ decreases progressively, the corresponding parameters become less sensitive to specific tasks or scenarios. This suggests that the layer has acquired detection meta-knowledge, making it a candidate for learngene extraction. To quantify this behavior, gradient significance is defined as the proportion of parameters in a layer with large gradients:
\begin{equation}  
    \rho^l_k = \frac{1}{\left|\mathbf{\Theta}_\mathrm{col}^l\right|} \sum_{i=1}^{\left|\mathbf{\Theta}_\mathrm{col}^l\right|} \Phi(g_{i,k}^l > \tau),
    \label{eq: gradient significance}
\end{equation}
where $\tau$ is a threshold, and $\Phi$ is an indicator function that returns 1 if $g_{i,k}^l > \tau$ and 0 otherwise. The sequence ${\rho^l_k}_{k=1,\dots,K}$ reflects the gradient magnitude trends within the $l$-th layer across tasks. If $\rho^l_k$ consistently decreases as $k$ increases, the corresponding layer is identified as a learngene candidate, ensuring that only meta-knowledge is retained for transfer.

Once the learngene $\mathbf{\Theta}_\mathrm{lg}$ consisting of $M$ layers is extracted, it is expanded by integrating randomly initialized layers $\mathbf{\Theta}_\mathrm{exp}$. This expansion ensures that the individual model inherits essential meta-knowledge while maintaining flexibility, enabling adaptation to diverse task-specific variations. To balance knowledge retention and adaptation, the individual model’s loss function is modified with L2 regularization:
\begin{equation}
    \mathcal{L}_\mathrm{ind} = \mathcal{L}(\mathbf{\Theta_\mathrm{ind}};\mathcal{D}_{K+1}) + \frac{\lambda}{2} \sum_{m=1}^{M} \left\| {\mathbf{\Theta}_\mathrm{lg}^m}' - \mathbf{\Theta}^m_\mathrm{lg} \right\|_2^2,
    \label{eq: loss ind}
\end{equation}
where $\mathbf{\Theta_\mathrm{ind}}=\{\mathbf{\Theta}_\mathrm{lg},\mathbf{\Theta}_\mathrm{exp}\}$ represents the individual model parameters, and $\mathcal{D}_{K+1}$ denotes the new task. Additionally, $\mathbf{\Theta}_\mathrm{lg}'$ refers to the learngene parameters after adaptation to $\mathcal{D}_{K+1}$. The regularization weight $\lambda$ controls the extent of learngene parameter updates, preserving meta-knowledge while allowing efficient adaptation to new tasks.

\subsection{Efficient Deployment of Detectors Using Learngene}
\label{subsec: Deployment pattern}
\begin{figure*}[!t]
    \centering
    \includegraphics[width=0.8\textwidth]{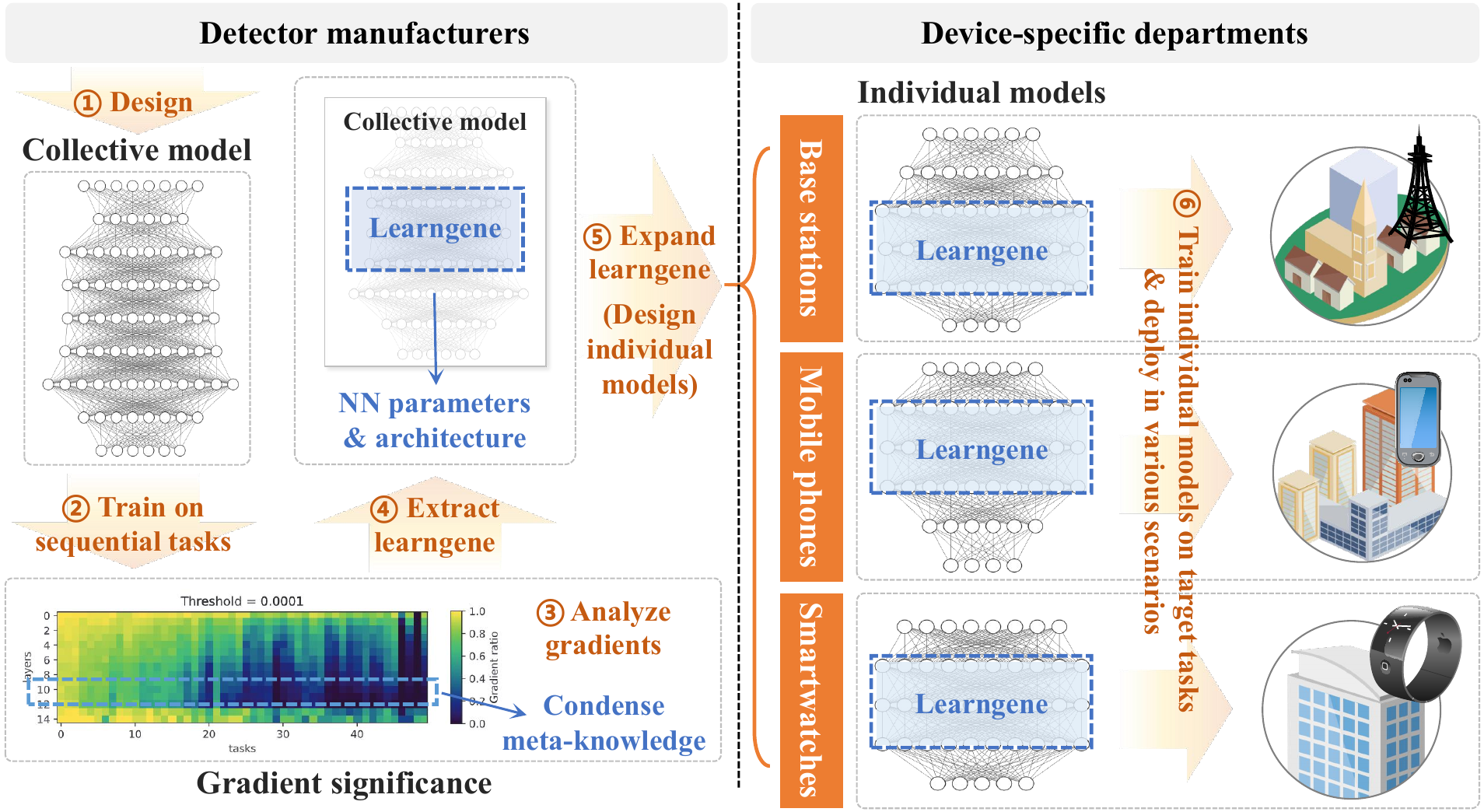}
    \caption{Deployment pattern of DL-based detectors with learngene.}
    \label{fig: deployment pattern}
\end{figure*}

The efficient deployment of DL-based detectors using learngene follows a structured two-phase process: learngene development at large detector manufacturers and learngene utilization at device-specific departments, as illustrated in Fig. \ref{fig: deployment pattern}. This framework leverages the computational capabilities and diverse CSI datasets available to manufacturers, enabling them to train collective models that learn generalized detection knowledge.

The first phase involves designing a large-scale collective detector that undergoes sequential training on various datasets to extract meta-knowledge. The learngene, a compact NN segment encapsulating detection meta-knowledge, is identified based on gradient significance analysis. Once extracted, the learngene is distributed to device-specific departments, enabling efficient adaptation without requiring access to the entire collective model.

In the second phase, individual detectors are designed using the extracted learngene to accommodate various devices such as BSs, mobile phones, and smartwatches. These lightweight individual models undergo fine-tuning with task-specific datasets before being deployed in real-world scenarios. Since the learngene serves as an effective initialization, device-specific departments only require minimal retraining, significantly reducing computational and resource costs.

This learngene-based deployment framework provides several key advantages:
\begin{itemize}
    \item \textbf{Reduced Deployment Costs}. By encapsulating detection meta-knowledge, learngene facilitates knowledge reuse across diverse scenarios and devices, minimizing the need for extensive retraining. This leads to significant cost reductions in training and deployment.

    \item \textbf{Flexibility Across Devices}. As a compact NN segment, learngene allows flexible customization of individual detector architectures, ensuring adaptability to devices with varying computational resources and constraints.
    
    \item \textbf{Data Privacy and Intellectual Property Protection}. The separation of learngene development and utilization eliminates the need for data sharing between manufacturers and device-specific departments, preserving CSI privacy. Additionally, distributing only the learngene rather than full models safeguards the intellectual property of detector designs.
\end{itemize}

\section{Experiments}
\label{sec: Simulation results}
In this section, we first describe the simulation settings, including key parameters for CSI generation, comparison schemes, and training configurations. We then present the experimental results, demonstrating the effectiveness of learngene selection, its rapid convergence, performance improvements, and scalability. Additionally, we analyze the reduction in computational resource requirements and discuss the rationale behind learngene extraction across different tasks.

\subsection{Simulation Settings}
\subsubsection{CSI Generation}

\begin{figure}[!t]
    \centering
    \includegraphics[width=0.43\textwidth]{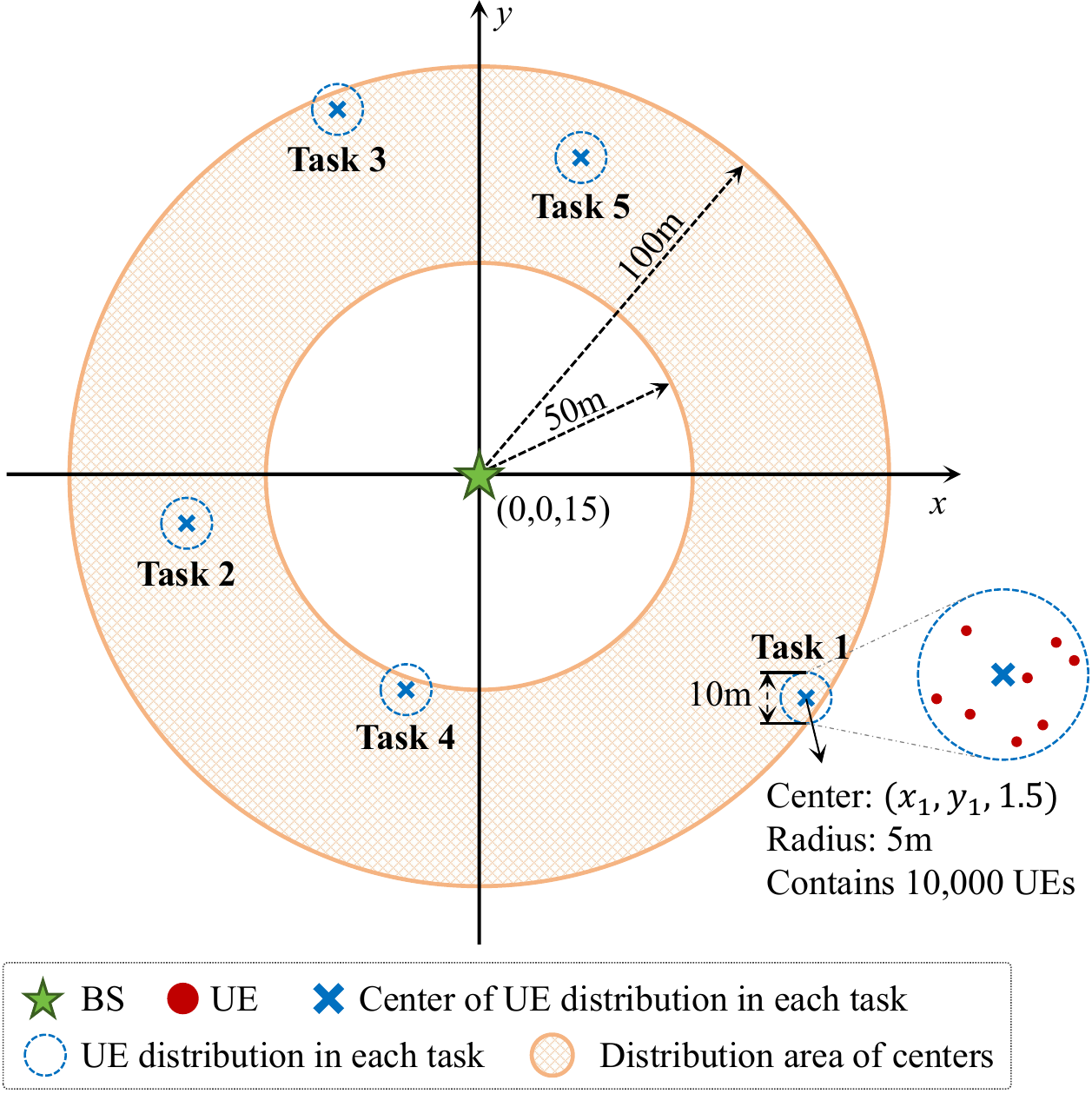}
    \caption{Illustration of the BS and UEs distribution during CSI generation, with only a subset of tasks and UEs displayed for clarity.}
    \label{fig: CSI generation}
    \vspace{-0.3cm}
\end{figure}

The CSI datasets are generated using QuaDRiGa \cite{jaeckel2014quadriga}, a channel model based on scatterer distribution. The simulation follows the 3GPP 38.901 urban-microcell (UMi) NLoS scenario. The number of antennas at the UE and BS is set to $N_\mathrm{t} = 8$ and $N_\mathrm{r} = 32$, respectively.
To simulate diverse communication conditions, the BS is fixed at $(0, 0, 15)$ meters, while 55 distinct UE distributions with varying scatterer configurations are generated, each considered an independent task, as illustrated in Fig. \ref{fig: CSI generation}.
To ensure task differentiation, UEs in each task are randomly positioned within a circular area of radius 5 meters, centered at $(x, y, 1.5)$ meters, where $(x, y)$ are randomly assigned within concentric circles of radii ranging from 50 to 100 meters.
For each task, 10,000 UEs and their corresponding CSI samples are randomly generated and split into: 8,100 samples for training, 900 samples for validation, and 1,000 samples for testing. The detailed simulation parameters are listed in Table \ref{tab: CSI generation}.

\begin{table}[!t]
\caption{Simulation settings for CSI generation.}
\centering
\renewcommand{\arraystretch}{1.5}
\begin{tabular}{p{3.5cm} p{4.5cm}} % {l l}
\toprule
\textbf{Parameter} & \textbf{Value} \\
\midrule
Carrier frequency & 6 GHz \\
Number of antennas (BS, UE)  & (32, 8) \\ 
Channel model & 3GPP 38.901 UMi NLoS \\
Number of tasks & 55 \\
BS position (m) & (0, 0, 15) \\
UE distribution in each task & Circular area (radius: 5 m)  \\
Center of UE distribution in each task (m) & \raisebox{-0.5\height}{$(x, y, 1.5), 50^2 < x^2 + y^2 < 100^2$} \\
UEs per task & 10,000 \\ 
Train/Val/Test split &  81\% / 9\% / 10\% \\
\bottomrule
\end{tabular}
\label{tab: CSI generation}
\end{table}

\subsubsection{Comparison Schemes}

Three schemes are considered for performance comparison: train-from-scratch, transfer learning, and learngene-based training. The specifics of each scheme are outlined below. 
\begin{itemize}
    \item \textbf{Train-from-scratch:} The target task is trained directly on a randomly initialized model without pre-existing meta-knowledge. The results may vary due to different initializations.
    
    \item \textbf{Transfer learning:} This scheme follows a pre-training and fine-tuning approach. The model is first pre-trained on a randomly selected source task from the training dataset and then fine-tuned on the target task, allowing it to acquire limited meta-knowledge beneficial for learning.
    
    \item \textbf{Learngene:} As discussed in Sections \ref{subsec: Learngene} and \ref{subsec: Training strategies}, the model undergoes sequential training on multiple tasks to extract the learngene based on gradient trends. The individual model is partially initialized with the learngene and further trained on the target task, leveraging the meta-knowledge accumulated by the collective model.
    
\end{itemize}

For simplicity, we refer to the train-from-scratch model, the fine-tuned transfer learning model, and the individual learngene-based model as target models. All target models share the same NN architecture, as illustrated in Fig. \ref{fig: NNarchitecture}. The primary performance metric for evaluating SDNet is the SER.

\subsubsection{Training Configuration}

\begin{table}[!t]
\centering
\caption{Training configuration for the learngene scheme.}
\label{tab: training configuration}
\setlength{\tabcolsep}{2pt}
\renewcommand{\arraystretch}{1.3}
\begin{tabular}{p{3.5cm} p{2.4cm} p{2.4cm}}
\toprule
\textbf{Parameter} & \textbf{Collective Model} & \textbf{Individual Model} \\
\midrule
\multicolumn{3}{l}{\textbf{NN Parameters}} \\
\midrule
Training tasks & 50 (sequential) & Unseen task \\
Layers & $N_\mathrm{conv}^\mathrm{col} = 12$ & $N_\mathrm{conv}^\mathrm{ind} = 8$ \\
Loss function & MSE & MSE + L2 (\ref{eq: loss ind}), $\lambda=2e^{-15}$ \\
Learning rate \& Optimizer & \multicolumn{2}{c}{0.001, Adam} \\
Batch size & \multicolumn{2}{c}{500} \\
Epochs & 50 per task & 200 \\
Learngene selection & \multicolumn{2}{c}{Gradient analysis (\ref{eq: gradient significance}), $\tau=1e^{-4}$} \\
\midrule
\multicolumn{3}{l}{\textbf{Communication Parameters}} \\
\midrule
SNR & \multicolumn{2}{c}{25 dB} \\
Number of pilots & \multicolumn{2}{c}{$N_\mathrm{p} = 16$} \\
\bottomrule
\end{tabular}
\end{table}

The implementation is based on Keras and TensorFlow. Due to the sequential training process of the collective model, 55 scenarios are generated, with 50 used for training the collective model and 5 reserved as target tasks for individual model training. To validate the effectiveness of the collective-individual framework and the learngene approach, a series of experiments are conducted on SDNet. Notably, the collective model has a larger architecture than the individual model, with $N_\mathrm{conv}^\mathrm{col} = 12$ and $N_\mathrm{conv}^\mathrm{ind} = 8$. The detailed NN and communication parameters are listed in Table \ref{tab: training configuration}.

Furthermore, the location of the learngene within the collective model and its inheritance into the individual model under different expansion strategies are described as follows. In the collective model, with the architecture shown in Fig. \ref{fig: NNarchitecture}, the alternative learngene consists of $N_\mathrm{conv}^\mathrm{col} = 12$ layers, equally divided into three segments: top, middle, and bottom, each containing four layers. In the individual model, the inheritable positions involve $N_\mathrm{conv}^\mathrm{ind} = 8$ layers, partitioned into the top (1st to 4th layers), middle (3rd to 6th layers), and bottom (5th to 8th layers).

\subsection{Learngene Selection Effectiveness}

\begin{figure}[!t]
    \centering
    \includegraphics[width=0.5\textwidth]{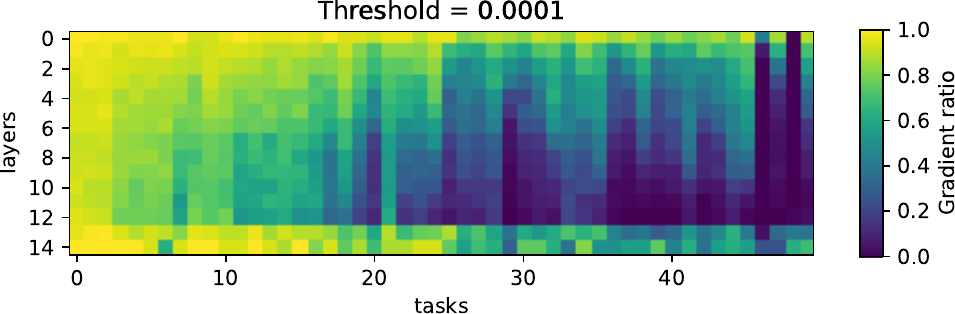}
    \caption{Gradient trends of the collective model during training on 50 sequential tasks at SNR = 25 dB.} 
    \label{fig: gradient trends}
\end{figure}

To identify an effective learngene, the collective model is sequentially trained using the parameters in Table \ref{tab: training configuration}. The gradient trends of the Conv layers and the FC layer in the collective model are shown in Fig. \ref{fig: gradient trends}, where the x-axis represents the task ID and the y-axis denotes the layer ID. The gradient significance of the final epoch in each task is recorded, excluding BN layers due to their limited trainable parameters. Additionally, the first and last Conv layers, as well as the FC layer, are excluded due to incompatible dimensions, leaving $N_\mathrm{conv} = 12$ significant layers, numbered 1 to 12 in Fig. \ref{fig: NNarchitecture}. Based on a predefined threshold, the final four Conv layers (IDs 9--12) exhibit the lowest gradient significance, suggesting that essential detection meta-knowledge is concentrated within these layers. Thus, they are selected as learngene.

During simulations, inheriting BN layers does not improve individual model training compared to inheriting only Conv layers. This is likely because BN layer parameters depend on the training dataset \cite{ioffe2015batch}. When the CSI distribution changes due to scenario variations, BN layers may degrade performance. Consequently, only Conv layers are inherited in the individual model, while BN layers are excluded.

\begin{figure}[!t]
    \centering
    \includegraphics[width=0.45\textwidth]{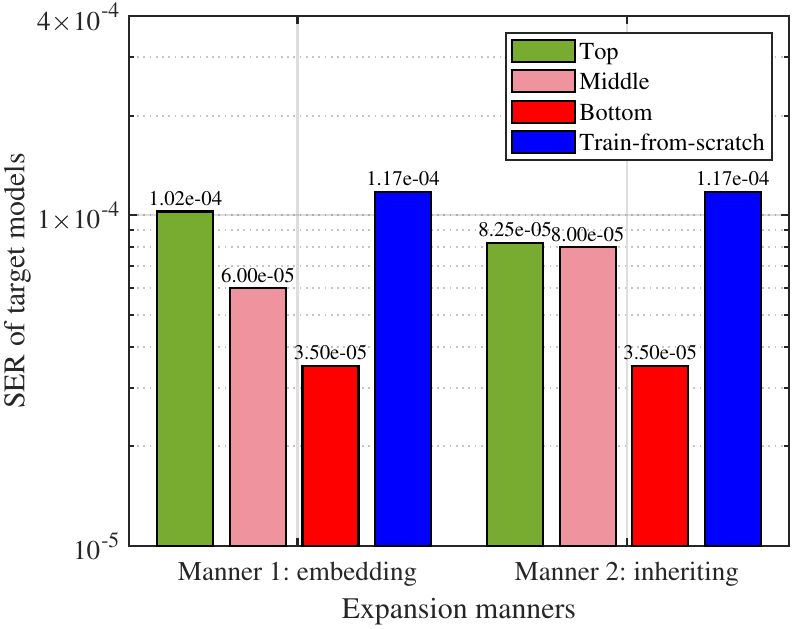}
    \caption{SER comparison between train-from-scratch and different learngene expansion strategies.} 
    \label{fig: SER scratch manners}
\end{figure}

\begin{figure}[!t]
    \centering
    \subfloat[Embedding.]{\includegraphics[width=0.45\textwidth]{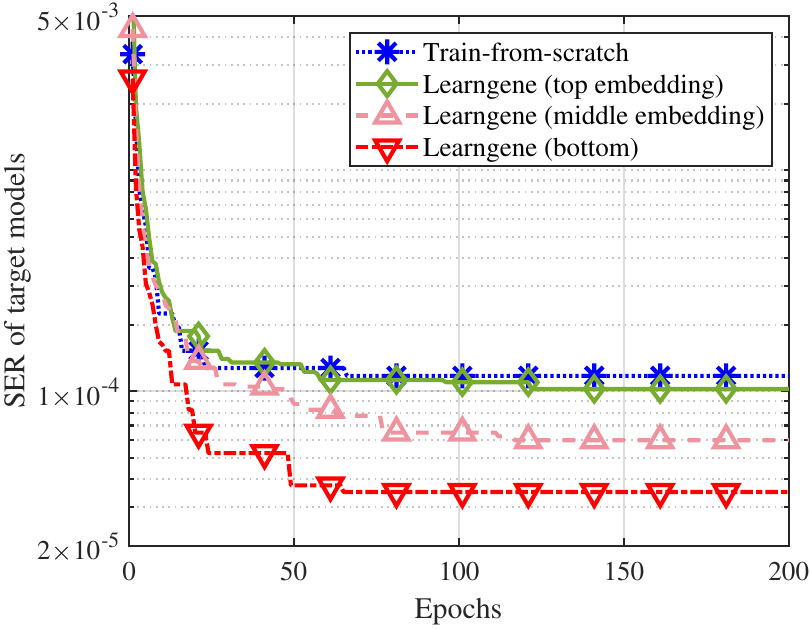}
    \label{fig: convergence rate-embed}}\\
    \subfloat[Inheriting.]{\includegraphics[width=0.45\textwidth]{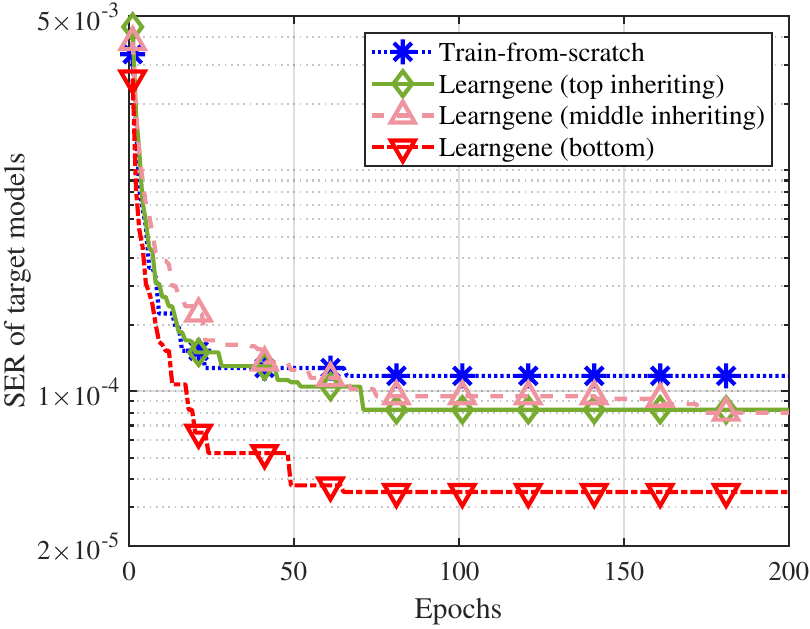}
    \label{fig: convergence rate-inherit}}
    \caption{Comparison of convergence rates between train-from-scratch and different learngene expansion strategies.} 
    \vspace{-0.4cm}
\end{figure}

The embedding location of the learngene in the individual model is critical for maximizing the meta-knowledge effect. We first evaluate different embedding strategies, as illustrated in Fig. \ref{fig: expansion manners}. The left half of Fig. \ref{fig: SER scratch manners} shows that the SER for the three embedding strategies and train-from-scratch are 1.02e-04 (top), 6.00e-05 (middle), 3.50e-05 (bottom), and 1.17e-04, respectively. The bottom embedding strategy achieves the best performance, and all three learngene strategies outperform train-from-scratch. Regarding convergence rate (Fig. \ref{fig: convergence rate-embed}), the bottom embedding scheme converges the fastest, followed by the middle embedding scheme, while train-from-scratch exhibits a convergence rate similar to the top embedding scheme. These results indicate that embedding the learngene at the bottom of SDNet yields the best performance.

To further evaluate the extracted learngene, we compare different inheriting strategies in Fig. \ref{fig: expansion manners}. Notably, bottom embedding and bottom inheriting correspond to the same strategy. The right half of Fig. \ref{fig: SER scratch manners} shows that the SER for top and middle inheriting strategies is 8.25e-05 and 8.00e-05, respectively---both at least twice as high as the bottom inheriting strategy. Additionally, the convergence rate analysis in Fig. \ref{fig: convergence rate-inherit} confirms that the bottom inheriting strategy converges the fastest. These findings align with our gradient significance analysis.

In conclusion, within SDNet, the most effective learngene consists of the four bottom Conv layers in the collective model. Embedding or inheriting the learngene at the bottom of the individual model maximizes meta-knowledge utilization, accelerates convergence, and enhances performance compared to other expansion strategies.

\subsection{Performance Evaluation of SDNet with Learngene}
\subsubsection{Fast Learning of Target Tasks}
\begin{figure}[!t]
    \centering
    \includegraphics[width=0.45\textwidth]{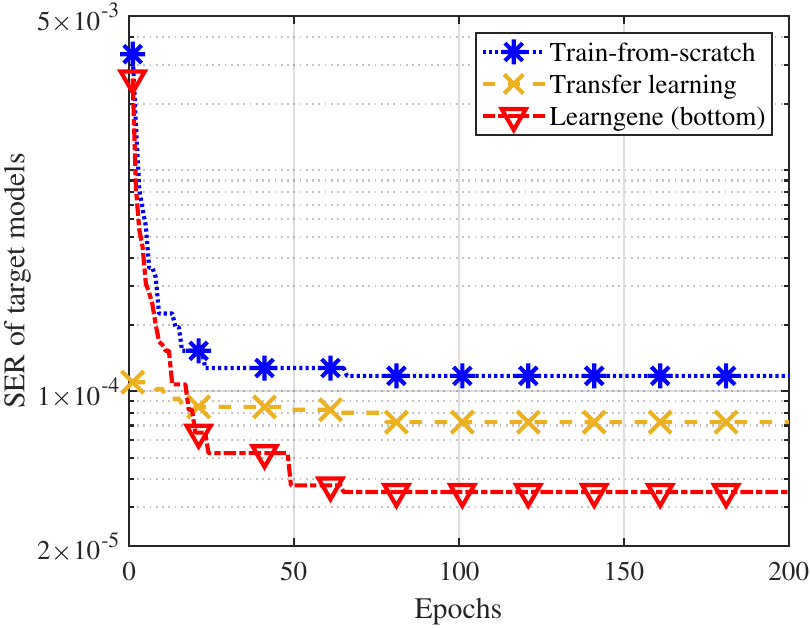} 
    \caption{Comparison of the convergence rate among train-from-scratch, transfer learning, and learngene.}
    \label{fig: convergence rate}
    \vspace{-0.4cm}
\end{figure}

As shown in Fig. \ref{fig: convergence rate}, we evaluate the convergence rate of learngene with a bottom-inheriting expansion strategy and compare it to train-from-scratch and transfer learning. The relationship between epochs and the SER of the target models is depicted in the figure. The train-from-scratch approach exhibits the slowest convergence, maintaining the highest SER throughout training. While transfer learning initially converges the fastest within the first 30 epochs, its convergence rate subsequently slows, falling behind learngene. Although learngene starts with a slower convergence rate, it demonstrates a significant advantage in later training stages, ultimately achieving the best overall performance.

This phenomenon arises due to the absence of meta-knowledge accumulation in the train-from-scratch approach, which forces the NN to learn both CSI features and the signal detection process from scratch. In contrast, transfer learning benefits from pre-training but is limited by the restricted diversity of the CSI dataset. Additionally, the pre-trained model is optimized for the source dataset rather than the target task, leading to suboptimal generalization.

In the learngene approach, the collective model evolves progressively through sequential training. The extracted learngene unit, enriched with detection meta-knowledge, is used to initialize the individual model, enabling faster adaptation and superior performance. While transfer learning achieves faster convergence in the early training phase, it requires storing the full NN architecture, whereas learngene only retains four Conv layers, reducing storage complexity. A detailed complexity analysis will be provided in the following section.

\subsubsection{SER Performance Under Different SNRs}
\begin{figure}[!t]
    \centering
    \includegraphics[width=0.45\textwidth]{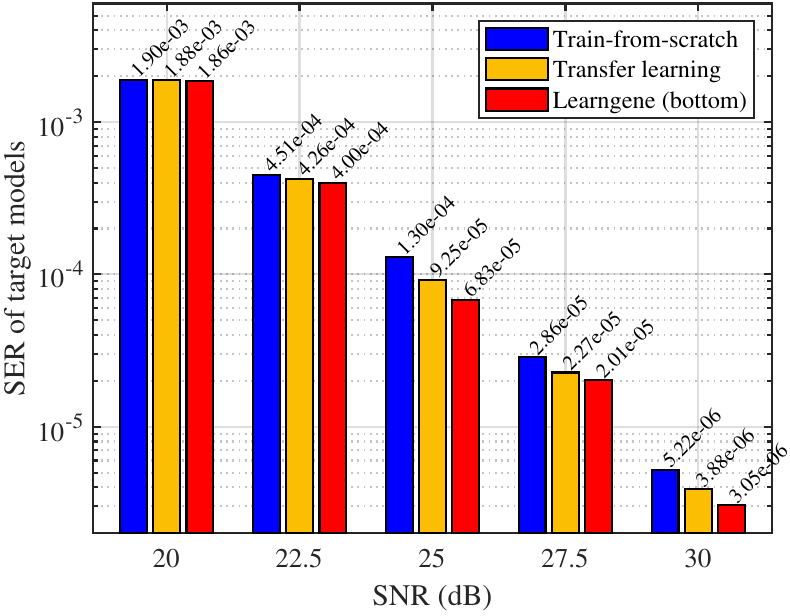}
    \caption{SER comparison of different schemes under varying SNRs.} 
    \label{fig: different SNR}
\end{figure}

Fig. \ref{fig: different SNR} illustrates the SER performance of learngene and two benchmark schemes across varying SNRs, ranging from 20 dB to 30 dB in 2.5 dB increments. After training and saving the target models for all three schemes, each model is evaluated on the test dataset to obtain the SER. As shown in the figure, SER decreases with increasing SNR for all schemes. Additionally, learngene consistently outperforms both benchmarks across all tested SNRs, achieving a 1 dB improvement over train-from-scratch and a 0.6 dB improvement over transfer learning. These results demonstrate the effectiveness of integrating SDNet with learngene across a broad range of SNR conditions, with performance gains becoming more pronounced at higher SNRs. This highlights learngene as a promising solution for signal detection in wireless communication systems. 

\subsubsection{Scalability to Different Antenna Configurations}

\begin{figure}[!t]
    \centering
    \includegraphics[width=0.45\textwidth]{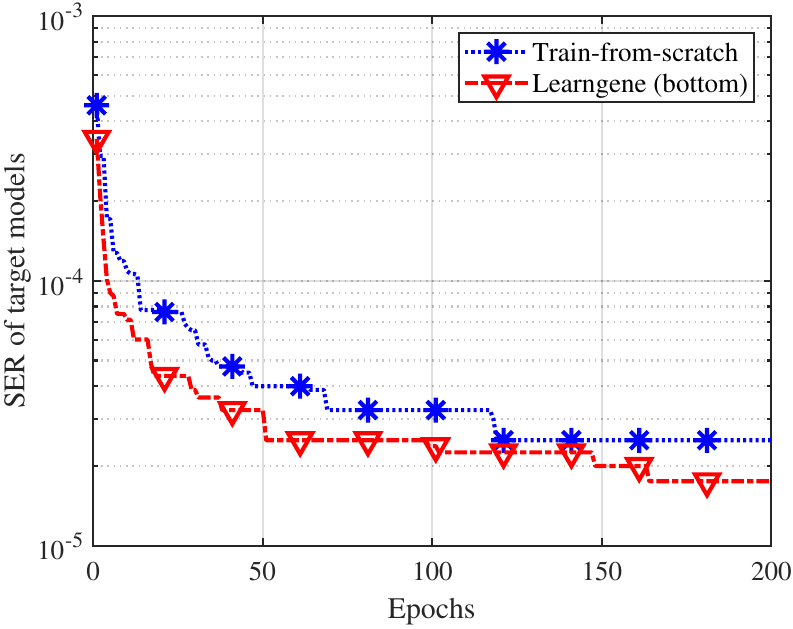} 
    \caption{SER of models with $N_\mathrm{t}=4, N_\mathrm{r}=32$. The learngene is extracted from the collective model trained with the parameters in Table \ref{tab: training configuration}.}
    \label{fig: antenna}
    \vspace{-0.4cm}
\end{figure}

To evaluate the scalability of learngene, we examine its performance under varying antenna configurations. The learngene used in this experiment is identical to that in previous trials, where $N_\mathrm{t}=8$ and $N_\mathrm{r}=32$. The system is tested with a different antenna configuration, $(N_\mathrm{t}, N_\mathrm{r})=(4,32)$. To ensure learngene reusability, an upsampling layer is added to the NN framework. The target models for both learngene and train-from-scratch remain identical, while the previous pre-trained model becomes inapplicable due to the configuration change. As shown in Fig. \ref{fig: antenna}, the learngene approach achieves faster adaptation to the target task compared to the train-from-scratch method, validating its scalability and adaptability to different scenarios and hardware configurations.

\subsection{Complexity Analysis}

\begin{table*}[!t]
\centering
\setlength{\tabcolsep}{7pt}
\renewcommand{\arraystretch}{1.5}
\caption{Comparison of computational complexity and transferred parameter ratios among the three schemes.}
\label{tab: complexity analysis}
\begin{tabular}{llcccc}
\toprule
\multicolumn{2}{l}{\textbf{Scheme}} & \textbf{Layers} & \textbf{Trainable Parameters} & \textbf{FLOPs (Conv + FC layers)} & \textbf{Transferred Parameters Ratio} \\
\midrule
\multicolumn{2}{l}{Train-from-scratch} & 10 Conv+BN, 1 FC & 21,627 & 9,917,440 & 0\% \\
\multicolumn{2}{l}{Transfer learning} & 10 Conv+BN, 1 FC & 21,627 & 9,917,440 & 100\% \\
\midrule
\multirow{3}{*}{Learngene} & Collective model & 14 Conv+BN, 1 FC & 24,027 & 14,709,760 & \multirow{3}{*}{\textbf{10.8\%}} \\
& Learngene unit & 4 Conv & 2,336 & 4,792,320 & \\
& Individual model & 10 Conv+BN, 1 FC & 21,627 & 9,917,440 & \\
\bottomrule
\end{tabular}
\end{table*}

Table \ref{tab: complexity analysis} summarizes the computational complexity and parameter transfer requirements for train-from-scratch, transfer learning, and learngene approaches. Specifically, it provides details on the layers included in each model, the number of trainable parameters, and the floating point operations (FLOPs) required, characterizing the computational complexity. Additionally, the transferred parameter ratio is computed as the number of transferred parameters divided by the total parameters of the target model.

As shown in the last column of Table \ref{tab: complexity analysis}, the train-from-scratch approach involves no transferred parameters, as all model parameters are randomly initialized. In contrast, transfer learning requires transferring 100\% of the pre-trained model's parameters (21,627 parameters) to the fine-tuned model, leading to significant storage overhead. Moreover, direct fine-tuning of a pre-trained model becomes impractical when deployment conditions change, necessitating updates to the pre-trained model and incurring additional training and computational costs.
In contrast, learngene transfers only a subset of parameters from the collective model, significantly reducing storage and computational demands. The extracted learngene unit is tailored to specific resource constraints, making it adaptable to various devices. Furthermore, learngene is acquired through a one-time training process of the collective model, avoiding the need for retraining under different deployment scenarios.

By analyzing performance, convergence rate, storage requirements, and training overhead, learngene provides an optimal balance among these factors. Compared to train-from-scratch, SDNet with learngene requires only the extraction and transfer of a lightweight learngene unit, incurring minimal storage costs while achieving faster convergence and superior performance. Compared to transfer learning, learngene not only outperforms in SER and convergence speed but also substantially reduces both storage requirements and training overhead when transferring knowledge across heterogeneous NNs.

\subsection{Discussion}

\begin{figure}[!t]
    \centering
    \includegraphics[width=0.45\textwidth]{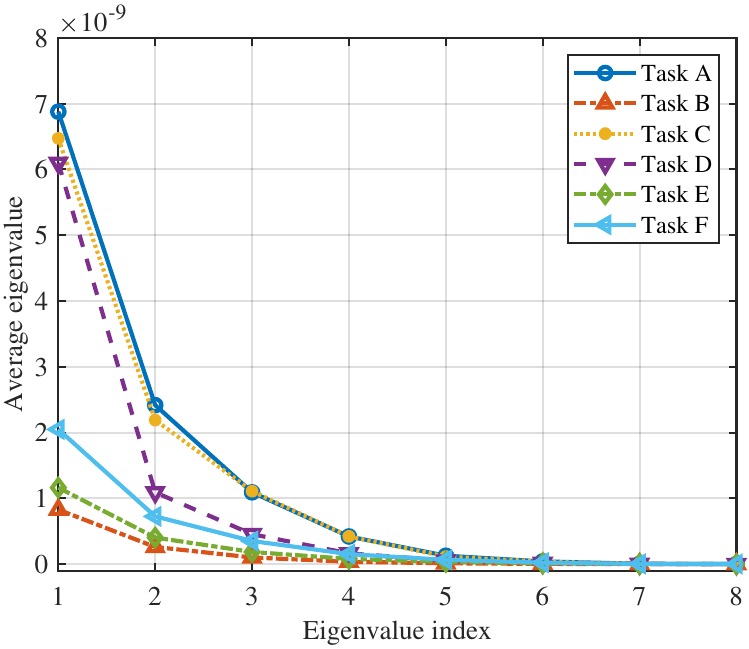}
    \caption{Eigenvalue spectrum of $\bar{\mathbf{H}}^H\bar{\mathbf{H}}$ across different  scenarios.}
    \label{fig: eig-of-tasks}
    \vspace{-0.4cm}
\end{figure}

To understand the rationale behind learngene extraction in different machine learning tasks, we compare the extraction process in image classification \cite{wang2022learngene} and denoising tasks. Both CSI feedback in \cite{li2024learngene} and signal detection in this work share a fundamental goal: reconstructing cleaner representations from noisy observations. Consequently, they can be categorized as denoising tasks. While all tasks in this comparison utilize convolutional NNs (CNNs) and extract bottom layers as learngene, the motivations behind this selection differ. 

In image classification, the study in \cite{wang2022learngene} demonstrated via visualization that the learngene unit primarily captures semantic information. This aligns with the general understanding of CNNs, where top layers specialize in task-specific features, which are sensitive to category shifts and local textures, while bottom layers encode semantic-level knowledge. Since semantic information remains consistent across multiple classification tasks, the bottom layers are identified as learngene, encapsulating transferable meta-knowledge. 

In contrast, denoising tasks rely on statistical feature representations that suppress noise while preserving essential signals. In the context of signal detection, SDNet is designed to mitigate the error component $\bar{\mathbf{n}}_\mathrm{ZF}$ introduced by ZF  detection. As derived in (\ref{eq: xzf2})-(\ref{eq: variance}), $\bar{\mathbf{n}}_\mathrm{ZF}$ follows a colored Gaussian distribution. To characterize its statistical properties, SDNet primarily focuses on the eigenvalue spectrum of CSI, which provides a more comprehensive representation than a single scalar (such as variance). The eigenvalue spectrum reflects how $\bar{\mathbf{n}}_\mathrm{ZF}$ is distributed across different data streams or feature spaces. Fig. \ref{fig: eig-of-tasks} illustrates the eigenvalue spectrum of $\bar{\mathbf{H}}^H\bar{\mathbf{H}}$ for randomly selected tasks. Additionally, $\bar{\mathbf{n}}_\mathrm{ZF}$ exhibits spatial correlation, where the noise, after passing through ZF detection, transforms from independent components into a correlated structure across different data streams.

These observations reveal how learngene is extracted in signal detection. During sequential training, the top layers of the collective model primarily learn local noise textures from the input, capturing the energy distribution of $\bar{\mathbf{n}}_\mathrm{ZF}$ across different data streams. Meanwhile, the bottom layers integrate this local information into global representations, implicitly reconstructing the eigenvalue spectrum of the channel. This process enables the model to distinguish the original transmitted signal $\mathbf{x}$ from noise and distortion, thereby enhancing denoising performance.

Since CSI varies across different scenarios, the statistical properties of $\bar{\mathbf{n}}_\mathrm{ZF}$ also change, affecting noise correlation, local textures, and energy distributions. Consequently, the top layers of the model focus on task-specific knowledge, whereas the bottom layers learn generalizable noise characteristics that are shared across different tasks. Similar to semantic feature extraction in image classification, the bottom layers in signal detection extract meta-knowledge that remains consistent across tasks. This property makes the bottom layers an ideal choice for learngene extraction.

\section{Conclusion}
\label{sec: Conclusion}

This paper addresses the challenges of high training costs and limited generalization capabilities in deploying DL-based MIMO detectors by integrating the learngene technique. The proposed approach leverages detector manufacturers with extensive datasets and computational resources to train collective models, from which a reusable and lightweight learngene unit is extracted. This learngene unit can be seamlessly integrated into various detectors, enabling fast adaptation across multiple devices and deployment scenarios.
Compared to existing knowledge transfer techniques, learngene offers a flexible and efficient mechanism for transferring detection meta-knowledge across heterogeneous detectors, significantly improving training efficiency. Simulation results validate the effectiveness of the proposed approach, showing that detectors incorporating learngene achieve faster convergence and enhanced performance, with transferred parameters accounting for only 10.8\% of the individual model.

\bibliographystyle{IEEEtran}
\bibliography{reference.bib}
\clearpage

\end{document}